\documentclass[reqno, centertags]{amsart}
\usepackage{amssymb}
\usepackage{graphicx}
\newcommand{\bmu}{{\mbox{\boldmath$\mu$}}}
\newcommand{\bnu}{{\mbox{\boldmath$\nu$}}}
\newcommand{\bla}{{\mbox{\boldmath$\lambda$}}}
\theoremstyle{definition}
\newtheorem{example}{Example}[section]
\begin{document}

\title{A probabilistic approach
to Quantum Mechanics based on `tomograms'}

\author{Michele Caponigro}

\address{Physics Department, University of Camerino, I-62032 Camerino, Italy }

\email{michele.caponigro@unicam.it}

\author{Stefano Mancini}

\address{Physics Department, University of Camerino, I-62032 Camerino, Italy }

\email{stefano.mancini@unicam.it}

\author{Vladimir I. Man'ko}

\address{P. N. Lebedev Physical Institute, Leninskii Prospect 53, Moscow 117924, Russia }

\email{manko@sci.lebedev.ru}

\date{\today}

\begin{abstract}
It is usually believed that a picture of Quantum Mechanics in
terms of true probabilities cannot be given due to the uncertainty
relations. Here we discuss a  \emph{tomographic} approach to
quantum states that leads to a probability representation of
quantum states. This can be regarded as a classical-like
formulation of quantum mechanics which avoids the counterintuitive
concepts of wave function and density operator. The relevant
concepts of quantum mechanics are then reconsidered and the
epistemological implications of such approach discussed.
\end{abstract}

\maketitle

\section{Introduction}

Since the early days of quantum mechanics, we have been
forced to coexist with wave functions \cite{Sch26}, therefore with
complex probability amplitudes without worrying about
their lack of any
reasonable physical meaning. One should not ignore,
however, that the
wave-like properties of quantum objects still raise
conceptual problems on
whose solutions, a general consensus is far from
having been
reached~\cite{Sch35,Ein35}.
A possible way out of this difficulty has been
implicitly suggested by
Feynmann~\cite{Fey87}, who has shown that, by dropping
the assumption that the
probability for an event must always be nonnegative, one
can avoid the use
of probability amplitudes in quantum mechanics. This
proposal goes back to
the work by Wigner~\cite{Wig32}, who first introduced
nonpositive
pseudoprobabilities to represent quantum mechanics in
phase space, and to
the Moyal approach to quantum mechanics~\cite{Moy49}.

From a conceptual point of view, the elimination of the
waves from quantum
theory is in line with the procedure inaugurated by
Einstein with the
elimination of the aether in the theory of electromagnetism
\cite{Ein52}.  Then, quantum mechanics without wave
functions has been discussed by several authors~\cite{Str56}.

The predictive character of the wave function
led to a statistical description in terms of the
density operator \cite{Neu32}, which in turn
gave rise to speculations about possible analogies with classical
statistical mechanics \cite{Dav69}.
Along this line
the phase-space formulation of quantum
mechanics~\cite{Wig32,Hus40,Gla63}
provides a means of analyzing quantum-mechanical
systems while still
employing a classical framework.
However, this approach involves counterintuitive notions,
like quasi-probabilities,
dictated by the uncertainty relations.
In fact, as matter of quantum mechanics we cannot see physical
objects {\it as they are} since the overall back action of any
observation cannot be made less than Planck's constant $\hbar$.
Instead, we only see the various aspects of the physical objects, like
the wave or the particle aspects which depend on the particular kind
of observation. In this respect we are really like the prisoners in
Plato's famous parable \cite{Pla} who were chained in a cave and
forced to see only the shadows of the things outside but not the
things as they are. Can we infer the things from their shadows ?
{\it Tomography} is a method for building up a picture of a hidden object
from various observable projections (called {\it tomograms} or {\it
marginals}).

Recently, the problem of quantum state measurement,
initially posed
by Pauli \cite{Pau58}, received a lot of
attention \cite{Jmo97}.
The tomographic approach~\cite{Ber87,Vog89} to the
quantum state of a
system has allowed to establish a map between the
density operator (or any its
representation) and a set of marginal
distributions. The latter have all the
characteristics of classical
probabilities; they are nonnegative, measurable,
and normalized.
Based on this connection, a classical-like description
of a quantum system and its dynamics by
means of tomography could be
formulated \cite{Man97}, providing
a bridge between classical and quantum worlds in the same scenario.

Here, we would analyze such possibility more in details by also
considering implications and foundational aspects.
Essentially, our aim is to eliminate the hybrid
procedure of describing the
dynamical evolution of a system, which consists of
a first stage, where
the theory provides a deterministic evolution of the
wave function, followed
by a hand-made construction of the physically
meaningful probability
distributions. If the probabilistic nature of the
microscopic phenomena is
fundamental, and not simply due to our ignorance,
as in classical statistical
mechanics, why should it be impossible to describe
them in probabilistic
terms from the very beginning? On the other hand,
the language of
probability, suitably adapted to take into account
all the relevant
constraints, seems to be the only language capable
of expressing the
fundamental role of {\it chance} in nature~\cite{Kuh70}.

The paper is organized as follows.
In Sec.\ref{CST} we introduce the notion of classical states by means of tomograms.
Then, in Sec.\ref{QST} we switch to quantum states.
In Sec.\ref{Info} we discuss the completeness of a set of tomograms.
Sec.\ref{Ent} is devoted to the concept of entanglement and Sec.\ref{Meas} to the measurement problem.
Finally, Sec.\ref{QL} point out relations with quantum logic.
Sec.\ref{concl} is for conclusions.
%%%%%%%%%%%%%%%%%%%%%%%%%%%%%%%%%%%%%%%%%%%%%%%%%%%%%%%%%%%%%%%

\section{Classical States and Tomograms}
\label{CST}
We provide a picture of a classical state
by means of tomograms. This enables us to
also give a tomographic description of the system dynamics.

\subsection{Classical states}

The state of a system in classical statistical mechanics
is described by a probability distribution function
$f\left(q,\,p\right)$ in
the phase space $\{q,p\}\equiv {\mathbb{ R}}\times{\mathbb{ R}}$,
where $q$, $p$ are position and momentum coordinate respectively
(for the sake of simplicity we are considering a system
with one degree of freedom).
By definition the function $f\left(q,\,p\right)$
is semidefinite positive and normalized\footnote{
Throughout the paper if not specified the integral delimiters
must be intended from $-\infty$ to $+\infty$.}
$\int dqdp\,  f\left (q,\,p\right)=1$.
Furthermore,
\begin{equation}
\int f\left(q,\,p\right)\,dp\,,\qquad
\int f\left(q,\,p\right)\,dq\,,
\end{equation}
represent the {\it marginal} probability distributions for position
and momentum, respectively,
that is the probabilities one can get by measuring observable like
position or momentum.

Let us now consider a canonical transformation in the phase space
$\left(q,\,p\right)\to\left(x,\,y\right)$, described by
\begin{equation}
\left[
\begin{array}{c}
x
\\
y
\end{array}
\right]
={\mathcal M}
\left[
\begin{array}{c}
q
\\
p
\end{array}
\right]\,,
\qquad
{\mathcal M}=
\left[
\begin{array}{cc}
\mu & \nu
\\
\mu' & \nu'
\end{array}
\right]\,,\label{sym}
\end{equation}
where ${\mathcal M}$ is a real symplectic matrix, that is
${\mathcal M}{\mathcal S}{\mathcal M}^{T}={\mathcal I}$ with
${\mathcal S}=\left[\begin{array}{cc} 0&1\\ -1&0 \end{array}\right]$
and ${\mathcal I}$ the $2\times 2$ identity matrix.
The marginal related to the observable $x$ can be written as
\begin{equation}
w(x)=\int dy\,f\left(q\left(x,\,y\right),\,
p\left(x,\,y\right)\right)=
\int dqdp\,  f\left(q,\,p\right)\,\delta
\left (x(q,\,p)-x\right )\,.
\label{csm6}
\end{equation}
Therefore, the marginal $w(x)$ is a real nonnegative, measurable
function which is also normalized
\begin{equation}\label{csm9}
\int dx\, w\left (x\right )=\int  dqdp\,f\left( q,\,p\right )=1\,.
\end{equation}
As a consequence, the function $w$ represent a true probability for the
stochastic variable $x$, while it
parametrically depends on $\mu$, $\nu$.
Thus, the notation $w(x;\mu,\nu)$ would be more appropriate.

The variable $x\left (q,\,p\right )$ can be considered from two equivalent
points of view. It can be interpreted as a canonically transformed position
which is a linear combination of position and momentum in a fixed reference
frame in the phase space of the system. Another equivalent interpretation of
$x\left (q,\,p\right )$ is that it is a
position of the system measured in the rotated and scaled reference frame in
the classical phase space of the system.

We use the second interpretation, according to which the real parameters
$\mu $ and $\nu $ determine the reference frame in the phase space of the
system in which the position is measured. For the position in the transformed
reference frame, we get from Eq.(\ref{csm6}) the tomography map
\begin{equation}\label{csm11}
w\left (x;\,\mu ,\,\nu \right )=\frac {1}{2\,\pi}\int dq\,dp\,dk\, e^{-ik(x-\mu q-\nu p)}
\,f\left( q,\,p\right)\,.
\end{equation}
Eq.({\ref{csm11}) has the inverse \cite{Man99}
\begin{equation}\label{csm22}
f\left (q,\,p\right)=\frac {1}{4\,\pi^2}\int dx\,d\mu \,d\nu\,
w\left (x;\,\mu ,\,\nu \right)\,\exp \left [i\left (x-\mu q-\nu p\right )
\right ] \,.
\end{equation}

{\footnotesize \begin{example}
The partial case of the canonical transform (\ref{sym}) is a rotation in
the phase space for which
\begin{equation}\label{csm13}
{\mathcal M}=\left[\begin{array}{cc}
\cos \,\theta &\sin \,\theta
\\
-\sin \,\theta &\cos \,\theta
\end{array}
\right]\,.
\end{equation}
By introducing the notation for the marginal distribution of the
rotated position
$w\left (x;\,\theta \right )$
we get, in view of (\ref{csm11}),
\begin{eqnarray}\label{csm17}
w\left (x;\,\theta \right )&=&\frac {1}{2\,\pi}\int dq\,dp\,dk\,
e^{-ik\left(x-q\cos \,\theta -p\sin \,\theta \right )}\,
f\left (q,\,p\right )\nonumber\\
&=&\int dy\,  f\left (x\cos \,\theta -
y\sin \,\theta ,\,x\sin \,\theta +y\cos \,\theta \right )\,.
\end{eqnarray}
which is a Radon transform \cite{Rad17}.
Its inverse reads
\begin{equation}
f\left (q,\,p\right)=\frac {1}{4\,\pi^2}\int dk\,|k|
\int_{0}^{\pi}d\theta\int dx\,
w\left (x;\,\theta \right)\,e^{ik(x-q\cos\theta-p\sin\theta)}\,.
\end{equation}
In classical statistical mechanics, the admissible marginal distributions
of Eq.(\ref{csm22}) always satisfy the
condition that $f\left (q,\,p\right )$ is a nonnegative function.
\end{example}
}

We have shown that instead of the distribution function
$f\left (q,\,p\right)$ the state of the classical system in
the framework of classical statistical mechanics can be represented by
the marginal distribution function $w\left (x;\,\mu,\,\nu \right )$
(intended as the entire set of marginals-probabilities on $x$ corresponding
to all possible values of parameters $\mu$ and $\nu$).
 Since the map
$f\left (q,\,p\right )\mapsto w\left (x;\,\mu ,\,\nu
\right )$
is invertible \cite{Man99}, the information contained in the distribution function
$f\left (q,\,p\right )$ is equivalent to the information
contained in the marginal distribution $w\left (x;\,\mu ,\,\nu \right ).$

{\footnotesize\begin{example}
If one introduces the distribution function
\begin{equation}\label{1*}
f\left(q,\,p\right)=\delta \left(q-q_0\right)\delta \left(p-p_0\right),
\end{equation}
which describes a particle with fixed position $q_0$ and momentum $p_0$,
the marginal distribution takes the form
\begin{equation}\label{2*}
w\left(x;\,\mu,\,\nu\right)=\delta \left(x-\mu q_0-\nu p_0\right)
\end{equation}
or
\begin{equation}\label{3*}
w\left(x;\,\theta \right)=\delta \left(x-q_0\cos \theta -p_0\sin
\theta\right).
\end{equation}
For classical statistical mechanics, the discussed tomography map connects
the positive distributions, and in this context our understanding of
the notion of the classical state for systems with fluctuations is unchanged.
\end{example}
}

\subsection{Classical dynamics}

The evolution equation for the classical distribution function $f(q,p,t)$ of a particle
with unit mass and potential $U(q)$, reads \cite{Man99}
\begin{equation}\label{csm23}
\partial_{t}  f+
p\,\partial_{q}  f-
\partial_{q} U(q)\,\partial_{p}  f=0\,,
\end{equation}
where $\partial_{\bullet}$ indicates the partial derivatives with respect
to the variable $\bullet$.

As consequence of the map between $f$ and $w$ developed in the
preceeding section, Eq. (\ref{csm23}) can
be rewritten for time dependent marginal distribution,
$w\left (x;\,\mu ,\,
\nu ,\,t\right)$, as
\begin{equation}\label{csm24}
\partial_{t} w-\mu \partial_{\nu} w-\partial_{q} U
\left (\widetilde q\,\right)\left [\nu \partial_{x} w\right ]
=0\,,
\end{equation}
where the argument of the function $\partial_{q}U$ is replaced by
the operator
$\widetilde q=-\partial_{x}^{-1}\partial_{\mu}$ with
$\partial_{\bullet}^{-1}\equiv\int\,d\,\bullet$.

{\footnotesize\begin{example}
For a harmonic oscillator with unit frequency and potential energy
$U(q)=q^2/2$, Eq.~(\ref{csm24}) becomes very simple, that is
\begin{equation}
\partial_{t} w-\mu \,\partial_{\nu} w
+\nu \,\partial_{\mu} w=0\,.
\end{equation}
Then, the initial distribution~(\ref{1*}) takes the form
\begin{equation}\label{1**}
f\left(q,\,p,\,t\right)=\delta \left(p-p_0\right)
\delta \left(q-tp-q_0\right)\,,
\end{equation}
and the initial marginal distribution~(\ref{2*}) becomes
\begin{equation}\label{2**}
w\left(x;\,\mu,\,\nu,\,t\right)=\delta
\left(x-\mu tp_0-\mu q_0-\nu p_0\right).
\end{equation}
\end{example}
}

%%%%%%%%%%%%%%%%%%%%%%%%%%%%%%%%%%%%%%%%%%%%%%%%%%%%%%%%%%%%%%%
\section{Quantum States and Tomograms}
\label{QST}

We now straightforward extend the previous approach to the
quantum domain.

\subsection{Quantum States}
In quantum mechanics, a representation of the state in
phase space $\{q,p\}\equiv{\mathbb{ R}}\times{\mathbb{ R}}$
cannot be given in terms of probability distributions
due to the uncertainty principle which forbids a joint measurement
of position and momentum operators $\hat q$, $\hat p$.
The state can be represented through a \emph{quasiprobability} function
like the Wigner function $W(q,p)$ \cite{Wig32}, which
plays the same role of the distribution function
$f(q,p)$ in the calssical domain.
Let us consider the symplectic transformation ${\mathcal M}$
acting on operators vector
\begin{equation}
\left[
\begin{array}{c}
{\hat x}
\\
{\hat y}
\end{array}
\right]
={\mathcal M}
\left[
\begin{array}{c}
{\hat q}
\\
{\hat p}
\end{array}
\right]\,.
\end{equation}
Then, the analogous of Eq.(\ref{csm11}) now reads
\begin{equation}\label{w}
w\left (x;\,\mu,\,\nu \right )=\frac{1}{(2\pi)^{2}}
\int dk\,dq\,dp\,\exp \left [-ik(x-\mu q-\nu
p)\right ]W(q,\,p)\,.
\end{equation}
We use the same notation as in the classical case.
Again, the physical meaning of the parameters $\mu $ and $\nu $ is that
they describe an ensemble of rotated and scaled reference frames
in which the position $\hat x$ is measured. For $\mu =\cos \,\theta $
and $\nu =\sin \,\theta ,$ the marginal distribution~(\ref{w}) is
the distribution for the homodyne-output variable used in optical
tomography~\cite{Vog89}. Formula~(\ref{w}) can be inverted and
the Wigner function of the state can be expressed in terms of the
marginal distribution~\cite{Man97}\,:
\begin{equation}\label{W}
W(q,\,p)=\frac {1}{2\pi }\int d\mu \,d\nu \,dx\, w\left (x;\,\mu ,\,\nu \right )
\exp \left [-i\left (\mu q+\nu p-x\right )\right ]\,.
\end{equation}
Since the Wigner function determines completely the quantum state of a system
and, on the other hand, this function itself is completely determined by the
marginal distribution, one can understand the notion of the quantum state in
terms of the classical marginal distribution for squeezed and rotated
quadrature.
So, we say that the quantum state is given if the position probability
distribution $w\left(x;\,\mu,\,\nu \right)$ in an ensemble of
rotated and squeezed reference frames in the classical phase space is
given.

The description of quantum states by the probability function gives
the possibility to formulate quantum mechanics without using the wave
function or density matrix. These ingredients of the quantum theory can
be considered as objects which are not mandatory ones, since
they are not directly accessible, but mere abstractions.
The marginal probability distribution function
$w\left(x;\,\mu,\,\nu\right)$,  which can be measured directly, replaces
the wave function in the new formulation of quantum mechanics. Since
the quantum mechanics formalism is reduced to the formalism of
classical probability theory, well-known results of the probability
theory can be used to get new results in quantum theory.
One can also introduce a notion of distance between quantum states
in terms of distance between proabilities \cite{DMM98}.

{\footnotesize\begin{example}
An example of marginal without classical counterpart is given by Schr\"odinger cat state \cite{Sch35} for which
\begin{eqnarray}\label{cat}
w(x;\theta)&=&\frac{1}{2}\exp[-(q_0^2+p_0^2)]
\left\{\omega_A(x;\theta)-\omega_B(x;\theta)-\omega_B^*(x;\theta)+\omega_A(-x;\theta)\right\}\\
\omega_A(x;\theta)&=&\frac{1}{\sqrt{\pi}}\exp\left[
-x^2-q_0^2\cos^2\theta-p_0^2\sin^2\theta\right.\nonumber\\&&\left.+2xq_0\cos\theta+2xp_0\sin\theta-2q_0p_0\sin\theta\cos\theta\right]\\
\omega_B(x;\theta)&=&\frac{1}{\sqrt{\pi}}\exp\left[
-x^2-q_0^2\cos^2\theta-p_0^2\sin^2\theta\right.\nonumber\\&&\left.-2ixq_0\sin\theta+2ixp_0\cos\theta-2q_0p_0\sin\theta\cos\theta\right]
\end{eqnarray}
\end{example}
}
From the tomographic approach it is clear that there are some classical marginals  (states) that do not allow quantum counterpart due to the uncertainty principle. An example is provided by Eq.(\ref{2*}). On the other hand it is well known that there are quantum states (marginals) that do not allow classical counterpart. An example is provided by Eq.(\ref{cat}). Thus we conclude that classical theory is not properly included into quantum theory. They are rather distinct with only a partial overlap.

Moreover, since some classical tomograms are not admissible in quantum mechanics. they cannot be obtained by the limit procedure $\hbar\to 0$ from the quantum ones.
Thus classical mechanics is not the $\hbar\to 0$ limit of quantum mechanics, as much as like the semigroup cannot give the group properties \cite{OVM04}.

%%%%%%%%%%%%%%%%%%%%%%%%%%%%%%%%%%%%%%%%%%%%%%%%%%%%%%%%%%%%%%%
\subsection{Quantum Dynamics}
The Wigner function $W(q,p,t)$ evolution is given by the following equation \cite{Gar}
\begin{eqnarray}
&&\partial_tW+p\partial_qW-\partial_qU(q)\partial_pW\nonumber\\
&&\qquad-\sum_{n=1}^{\infty}\frac{(i\hbar/2)^{2n}}{(2n+1)!}\partial^{2n+1}_qU(q)\partial^{2n+1}_pW=0
\label{Wqdyn}
\end{eqnarray}
while the evolution equation for tomograms is \cite{Man97}
\begin{eqnarray}
&&\partial_tw-\mu\partial_\nu w-\nu\partial_qU(q)\partial_x w\nonumber\\
&&\qquad-2\nu\sum_{n=1}^{\infty}\frac{(i\nu\hbar/2)^{2n}}{(2n+1)!}\partial^{2n+1}_qU(q)\partial^{2n+1}_x w=0\label{wqdyn}
\end{eqnarray}
Notice that the first three terms of Eq.(\ref{wqdyn}) give the $\hbar\to 0$ classical Boltzmann equation.

{\footnotesize\begin{example}
For a harmonic oscillator with unit frequency and potential energy
$U(q)=q^2/2$, Eq.~(\ref{wqdyn}) becomes very simple, that is
\begin{equation}
\partial_{t} w-\mu \,\partial_{\nu} w
+\nu \,\partial_{\mu} w=0\,.
\end{equation}
\end{example}
}
More generally, quadratic Hamiltonians give the same evolution equations for classical and quantum tomograms.

%%%%%%%%%%%%%%%%%%%%%%%%%%%%%%%%%%%%%%%%%%%%%%%%%%%%%%%%%%%%%%%

\section{Information completeness}
\label{Info}

The problem of how to achieve a kind of measurement that is ``complete'' in the sense that it can be used to infer information on all possible (also exclusive) observables dates back to Ref.\cite{Pru77}.

Obviously enough, no set of sharp observables can be informationally complete, while a set of (partially) non-commuting unsharp observables can be informationally complete \cite{BL}.
Actually the problem of determining minimal sets of informationally complete observables can be traced back to a group theoretical problem which is still unsolved in its generality \cite{Dar00}. Nonetheless we know that the set of rotated position observables leading to $w(x;\theta)$ is informationally complete, because there is a one-to-one correspondence between $w(x;\theta)$ and the quantum state  \cite{Vog89}.
Thus, we may formulate the notion of quantum state as follows.
We say that the quantum state is given if the position
probability distribution $w\left(x;\,\theta \right)$ is given
for all possible rotated reference frames, i.e. $\forall\theta\in[0,\pi]$.
This corresponds to sample the parameter $\theta$ over the set $[0,\pi]$ with uniform probability measure ${\mathcal Q}(\theta)d\theta\equiv(1/\pi)d\theta$.

What happen if we have partial knowledge about $w(x;\theta)$,
that is if we know only some of the tomograms because the parameter $\theta$ has been sampled with
a real distribution ${\mathcal P}$ instead of the ideal one ${\mathcal Q}$?
We can use the relative entropy between ${\mathcal Q}$ and ${\mathcal P}$
to measure the degree of completeness of information achieved by ${\mathcal P}$ \cite{WE}:
\begin{equation}\label{Hrel}
H({\mathcal P}\| {\mathcal Q})\equiv -\int_0^{\pi}d\theta\,{\mathcal P}(\theta)\,\log\frac{{\mathcal Q}(\theta)}{{\mathcal P}(\theta)}.
\end{equation}

The information contained in the marginal
distributions $w(x;\mu,\nu)$ is somehow overcomplete.
In fact, to determine the quantum state completely, it suffices to give the
function for arguments with the constraints
$\left(\mu^2+\nu^2=1\right)$
which corresponds to the scheme
$\mu=\cos \theta$, $\nu=\sin\theta$.
In particular, suppose to measure the position in reference frames sampled with a distribution ${\mathcal P}(\mu,\nu)$. If we write $\mu$ and $\nu$ in polar coordinate, we know that the minimal set of observables to measure only relies on the angular variable, hence we can consider
\begin{equation}
{\mathcal P}(\theta)=\int_{-\infty}^{+\infty}dr\, r\, {\mathcal P}(\mu=r\cos\theta,\nu=r\sin\theta),\,\qquad \theta\in[0,\pi].
\end{equation}
and use it in Eq.(\ref{Hrel}).
Obiously, if ${\mathcal P}$ is uniform over $\theta$ we get a complete information irrespective of the behavior along $r$.

That informationally complete measurements are relevant for
foundations of quantum mechanics as a kind of ``standard'' for a
purely probabilistic description has been pointed out also in
Ref.\cite{Fuc02}.

%%%%%%%%%%%%%%%%%%%%%%%%%%%%%%%%%%%%%%%%%%%%%%%%%%%%%%%%%%%

\section{Entanglement and Scaling transforms}
\label{Ent}

The generalization of the arguments of Sec.III to multipartite systems with $N$ degrees of freedom leads to tomograms of the kind $ w({\bf x};\bmu , \bnu)$, where ${\bf x},\bmu , \bnu$ are $N$-components vectors.
Let us define the tomographic dispersion matrix elements,
\begin{equation}
  \label{eq:tomo4}
  \sigma_{x_jx_k}(\bmu,\bnu) = \int d {\bf x} \, (x_j- \overline{x}_j ) (x_k - \overline{ x}_k )\,
  w({\bf x},\bmu , \bnu)
\end{equation}
where
\begin{equation}
\label{eq:tomo5}
\overline{x}_j   = \int d {\bf x} \, x_j \; w({\bf x}\,,\,\bmu \,,\, \bnu).
\end{equation}
Then, we can construct the dispersion matrix ${\mathcal V}$ using the following relations:
\begin{eqnarray}
  \label{eq:tomo6}
  {\mathcal V}_{jk} &=& \sigma_{x_jx_k}(\bmu={\bf 1}_{jk} \,,\, {\bnu} ={\bf 0} ),
  \qquad j,k=1,\ldots N \nonumber \\
   {\mathcal V}_{jk} &=& \sigma_{x_jx_j}({\bmu}={\bf 0} \,,\, {\bnu}={\bf 1}_{jk}),
     \qquad j,k=N+1,\ldots 2N\nonumber \\
 {\mathcal V}_{jk} &=& \sigma_{x_jx_k}({\bmu} ={\bf 1}_j \,,\, {\bnu} = {\bf 1}_k ),
 \qquad j=1,\ldots N;\;k=N+1,\ldots 2N
\end{eqnarray}
with ${\bf 1}_{jk}$ denoting the vector having $1$s in $j$th and $k$th components and zero elsewhere.

To study the separability (entanglement) we use the partial scaling transform to the system \cite{OVM05}. Starting from the tomogram of a state of the system $w({\bf x};\bmu,\bnu)$, we first verify that it satisfies the uncertainty principle by checking that ${\mathcal C} = {\mathcal V} + \frac{i}{2} {\mathcal R}\ge 0$, where ${\mathcal R}$ is the canonically invariant block diagonal matrix ${\rm diag}({\mathcal S},{\mathcal S},\ldots,{\mathcal S})$.

We can now perform an arbitrary scaling  described by the vector ${\bla} = (\lambda_1, \, \lambda_2 , \, \ldots \lambda_{2N})$ on the tomogram, that is $\mu_j\to\mu_j/\lambda_j$ and $\nu_j\to\nu_j/\lambda_j$. This implies a scaling on the dispersion matrix leading to
\begin{equation}
  \label{eq:gen1}
  {\mathcal C}^{{\bla}} = {\mathcal V}^{{\bla}}+ \frac{i}{2} {\mathcal R}
\end{equation}
where
\begin{equation}
  \label{eq:gen2}
  {\mathcal V}^{{\bla}} = {\mathcal D}_{{\bla}} {\mathcal V} {\mathcal D}_{{\bla}}
\end{equation}
with ${\mathcal D}_{{\bla}} \equiv {\mbox{diag}}(\lambda_1,\, \lambda_2, \, \ldots \lambda_{2N})$.
The $2N$ real parameters $\{\lambda_j \}$ parameterize the Abelian scaling semi-group and we require that
\[ |\lambda_1\lambda_2| \geq 1, \; |\lambda_3 \lambda_4| \geq 1 , \ldots , |\lambda_{2N-1}\lambda_{2N}| \geq 1.\]
The necessary condition for the separability of the state represented by the tomogram $w ({\bf x}, \bmu, \bnu)$ is then
\begin{equation}
  \label{eq:gen3}
  {\mathcal C}^{{\bla}} \geq 0
\end{equation}
for all allowed choices of ${\bla}$.

Out of the $2N$ scaling parameters we can always choose one pair, $(\lambda_{2k-1} ,\lambda_{2k})$ such that $|\lambda_{2k-1}\lambda_{k}| = 1$ using the freedom to choose an overall scale factor that does not affect the positivity of ${\mathcal C}^{{\bla}}$.
For two-mode systems, the choice $\lambda_1=\lambda_2=\lambda_3=1, \; \lambda_4 =\lambda^{-1}$ exhausts all the possibilities.

The scaling transformation is not a canonical transformation and can be thought of as an effective scaling of the Planck's constant. A separable state is not sensitive to such scalings applied to individual sub-systems. The entangled states, on the other hand, are sensitive to such scalings and in many cases this shows up by making ${\mathcal C}^{{\bla}}$ negative for certain choices of ${\bla}$.

Entanglement is synonymous of nonlocal correlations that can be contrasted with local ones through Bell-like inequalities involving only marginal distributions \cite{Evg}.

%%%%%%%%%%%%
\section{Quantum Measurements}
\label{Meas}

It is known \cite{WZ} that quantum  mechanics is problematic in
the sense that it is incomplete and needs the notion of a
classical device measuring quantum observables as an important
ingredient of the theory. Due to this, one accepts that there
exist two worlds: the classical one and the quantum one. In the
classical world, the measurements of classical observables are
produced by classical devices. In the framework of standard
theory, in the quantum world the measurements of quantum
observables are produced by classical devices, too. Due to this,
the theory of quantum measurements is considered as something very
specifically different from classical measurements.

It is psycologically accepted that to understand the physical meaning of a
measurement in the classical world is much easier than to understand the
physical meaning of analogous measurement in the quantum world.
Using the relations of the quantum states in the standard
representation and in the classical one (described by classical distributions),
one can conclude that complete information on a quantum state is obtained from
purely classical measurements of the position of a particle made
by classical devices in each reference frame of an ensemble of classical
reference frames, which are scaled and rotated in the classical phase space.

These measurements do not need any quantum language if we know how to
produce, in the classical world (using the notion of classical position and
momentum), reference frames in the classical phase space differing from each
other by rotation and scaling of the axis of the reference frame and how to
measure only the position of the particle from the viewpoint of these
different reference frames.

Thus, we avoid the paradox of the quantum world which requires for its
explanation measurements by a classical apparatus accepted in the framework of
standard treatment of quantum mechanics.
The problem of wave function collapse~\cite{WZ} reduces to the
problem of a reduction of the probability distribution which occurs as soon
as we ``pick'' a classical value of the classical random observable in the
classical framework.
This means that we ``solved'' the paradox of the  wave function collapse
reducing it to the problem of  standard measurement of a classical random
variable used in the probability theory.

Nevertheless, measurement on a reference frame affects the distributions on
the others (due to the underlying uncertainty principle), thus the nonlocal
character of QM is intrinsically present in a single system and emerges as
subtle correlations among distributions of different reference frames.

%%%%%%%%%%%%%%%%%%%%%%%%%%%%%%%%%%%%%%
\section{Relations with Logic}
\label{QL}

The developed approach provide a unified logical framework to
approach classical and quantum mechanics as summarized by the
following scheme.

\vspace{12 pt}
\setlength{\unitlength}{0.020in}

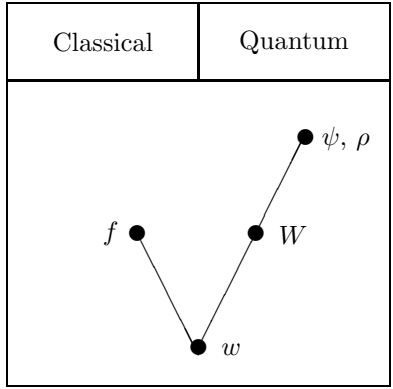
\begin{picture}(200,120)(-20,0)
\put(45,20){\framebox(100,100)}{
\put(80,58){\line(1,-2){15}}
\put(95,30){\line(1,2){28}}
\put(79,60){\circle*{4}}
\put(110,60){\circle*{4}}
\put(95,30){\circle*{4}}
\put(123,85){\circle*{4}}
\put(125,83){ $\psi$, $\rho$}
\put(101,28){$w$}
\put(70,58){$f$}
\put(116,57){$W$}
}
\put(45,100){\framebox(50,20){Classical}}
\put(95,100){\framebox(50,20){Quantum}}
\end{picture}

It is worth noticing that the orthodox quantum logic
(Birkhoff and von Neumann \cite{QLogic}) used \emph{projectors} as primitive elemants (propositions).
The great problem of quantum logic is that it is so general that it is not useful as model of quantum mechanics. We need more axioms than the assumptions made so as to build a bridge to Hilbert spaces.

Then, Algebraic Approach \cite{QLogic}
assumes \emph{observables} to be primitive and an algebra with natural physical assumtions of linearity, positivity and normalization is constructed. As a consequence there are states which have a place here though not in any classical theory. The aim was to prove the possibility of a quantum theory without Hilbert spaces and to show the possibility of deducing uncertainty principle from an abstract mathematical structure.

Also the tomographic approach does not need of Hilbert space construction. However, observables cannot have a prominent role over states because otherwise phenomenon like entanglement is precluded.

Then, one can consider the Convexity Approach \cite{QLogic}
where \emph{states} are taken as primitive. The principal concept is that of \emph{face}.
This approach has very important consequences for the problem of open systems.

Note that the impossibility of measuring the state of a single system also implies that quantum mechanics cannot only be a theory of states.
Then, a States-Observables Approach was developed by Mackey \cite{Mac}.
In this case both \emph{states and observables} are taken as primitives. Mackey developed a parallelism between quantum mechanics and classical probability calculus: observables are the random variables and the states are probability measures. It was tried to extend the discussion by inserting joint probability measures and conditional expectations: but the first exist only if the observables commute, and the second only if observables have discrete spectrum.
Our tomographic approach seems very much in line with Makey's logical construction and circumvent its problems, thus it could be useful to enforce it.

%%%%%%%%%%%%%%%%%%%%%%%%%%%%%%%%%%%%%%%%%%%
\section{Conclusions}
\label{concl}

We conclude that is possible to cast the standard
quantum mechanics into the form in which
only positive
probabilities are used to describe quantum states and
their evolution.
A possible
disadvantage of the approach proposed is a complicated evolution
equation~(\ref{wqdyn}), but,
perhaps, this is the price one ought to pay for the possibility of
describing quantum objects in terms
of classical probabilities.
Anyway, our arguments can constitute a step further
from the Bohr
position \cite{Bohr} about the inapplicability of
classical modes of
description in the quantum domain.
In fact,
while we belive that quantum mechanics is not classical physics
in disguise,
we retain (some) classical concepts still applicable against
counterintuitive notions like
complex statefunctions.

Moreover, our approach can be considered in line with what Wheeler argued: the origin of quantum mechanics structure is to be sought in a theory of observation and observers and meaning, then we would do well to focus our attention not on amplitudes but on quantities which are more directly observable.

However the approach distinguishes from recent information theoretic approaches to quantum mechanics \cite{Fuc02}, mostly based on epistemic attitude.
In such a cases quantum theory is reduced to simple postulates starting from information theoretic principles. This also leads to use probabilities with a `subjective' character.
Moreover, the epistemic point of view cannot be a realist view about the foundations of quantum theory.

The tomographic approach to quantum mechanics  is a classic approach, but the probability inferred is not similar with classic epistemic probability. We argue that our assumptions bring us to establish an ontological probability of the physical reality described by classic tomography (as stated in the introduction, if the probabilistic nature of the microscopic phenomena is
fundamental, and not simply due to our ignorance,
as in classical statistical mechanics, why should it be impossible to describe
them in probabilistic terms from the very beginning?). Thus our tomograms assume an ontological character and since they completely characterize the state of a system, also the latter assumes such a character. Due to Primas
\cite{PR}, the distinction between an ensemble interpretation and
an individual interpretation is characterized again in terms of
the distinction between an epistemic and an ontic interpretation
juxtaposing `description of our knowledge' to `description of
reality'. This distinction seems introducing in the first place
problems in the epistemic interpretation as being part of
psychology, not physics. Critically, it should be noted that
independently of its interpretation, a physical theory is a
representation of our knowledge, and therefore is always epistemic
(hence bringing us to a Kantian position).  Then, 
it is preferable to avoid the ontic versus epistemic
distinction in favour of distinction between individual object
versus ensemble.

As we seen, while quantum phenomena require a radical revision of our idea about physical reality, they do not prevent us from accepting a reasonable realistic individual interpretation. Quantum mechanics does not force us to give up realism, but it force us to distinguish carefully between \emph{potential} and \emph{actualized} properties. A popular working rule of pragmatic says that Òan observable has no value before a measurement.This is in contrast to the usual metaphysical commitment of classical mechanics that every observable has a value at all times.This commitment cannot be transferred to quantum mechanics since there is a theorem saying that for a full set of state of a
$C^*$-algebra, a hypothetical attribution of definite truth values to all elements requires that is commutative. However, instead of a positivistic renouncement we can adopt the intrinsic, internally consistent ontological interpretation that at every instant there is a maximal set of truth-definite observables, a truth definite observable possesses a value wheter we know this value or not, is at this stage of the theoretical discussion entirely irrelevant. This point of view corresponds exactly to the usual interpretation of classical point mechanics, where the ontological question of Òhaving a valueÓ is clearly separated from the entirely different question how to get empirically some information about this value.

Finally, the presented approach can be  extended to finite dimensional systems (like spin) \cite{OVM97} and relativistic systems \cite{PLA}.
An important analogy with methodology of special
relativity arises:~ It turns out that it is necessary to introduce a
consideration of events in the set of moving reference frames in space--time
in order to explain relativistic effects, and it is necessary to
introduce a consideration of events in the set of rotated and scaled
reference frames in the phase space in order to explain the nonrelativistic
quantum mechanics in terms of only classical concepts of classical fluctuation
theory. But these reference frames are the reference frames in the phase space
(not in space--time). Possibly, a combination of these two approaches  can be
generalized to give a classical description of relativistic
quantum mechanics.

%%%%%%%%%%%%%%%%%%%%%%%%%%%%%%%%%%%%%%%%%%%%%%%%%%%%%%%%%%%%%%%

%\section*{Acknowledgments}

%%%%%%%%%%%%%%%%%%%%%%%%%%%%%%%%%%%%%%%%%%%%%%%%%%%%%%%%%%%%%%%


\begin{thebibliography}{99}

\bibitem{Sch26}
E. Schr\"odinger, Ann. der Physik {\bf 79}, 489 (1926).

\bibitem{Sch35}
E. Schr\"odinger, Naturwissenshaften {\bf 49}, 53 (1935).

\bibitem{Ein35}
A. Einstein, B. Podolsky, and N. Rosen,
Phys. Rev. {\bf 47}, 777 (1935).

\bibitem{Fey87}
R. Feynmann, in {\it Quantum Implications},
B. J. Hiley and F. D. Peats~(Eds.),
(Routledge \& Kegan, London, 1987).

\bibitem{Wig32}
E. P. Wigner, Phys. Rev. {\bf 40}, 749 (1932);\\
E. P. Wigner, in {\it Perspectives in Quantum Theory},
W. Yourgrau and A. van der Merwe~(Eds.),
(Dover, New York, 1979).

\bibitem{Moy49}
J. E. Moyal, Proc. Cambridge Philos. Soc. {\bf 45}, 99 (1949).

\bibitem{Ein52}
A. Einstein, H. A. Lorentz, H. Minkowski and H. Weyl,
{\it The Principle of Relativity} (Dover, New York, 1952).

\bibitem{Str56}
R. L. Stratonovich, Zh. Eksp. Teor. Fiz. {\bf 31}, 1012 (1956);\\
L. Cohen, J. Math Phys. {\bf 7}, 781 (1966); {\bf 17}, 1863 (1976);\\
W. K. Wootters, Found. Phys. {\bf 16}, 391 (1986);\\
L. Wang and R. F. O'Connell, Found. Phys. {\bf 18}, 1023 (1988).

\bibitem{Neu32}
J. von Neumann, {\it Mathematische Grundlagen der Quantenmechanik},
(Springer, Berlin, 1932).

\bibitem{Dav69}
E. B. Davies,  Comm. Math. Phys.  {\bf 15}, 277 (1969);\\
{\it Quantum Theory of Open Systems}, (London, Academic Press, 1976).

\bibitem{Hus40}
K. Husimi, Proc. Phys. Math. Soc. Jpn {\bf 23}, 264 (1940).

\bibitem{Gla63}
R. J. Glauber, Phys. Rev. Lett. {\bf 10}, 84 (1963);\\
E. C. G. Sudarshan, Phys. Rev. Lett. {\bf 10}, 177 (1963).

\bibitem{Pla}
Plato, {\it Republic}, Book VII, in {\it The Loeb Classical Library},
{\bf L276}, Vol.VI (Harvard University Press, Cambridge, 1935).

\bibitem{Pau58}
W. Pauli, {\it Encyclopedia of Physics}, (Springer, Berlin, 1958),
Vol.5, p.17

\bibitem{Jmo97}
{\it Quantum State Preparation and Measurement},
J. Mod. Opt. {\bf 44}, (1997) special issue;\\
D. G. Welsch, W. Vogel and T. Opatrny,
Progress in Optics, {\bf XXXIX}, 63 (1999).

\bibitem{Ber87}
J. Bertrand and P. Bertrand, Found. Phys. {\bf 17}, 397 (1987).

\bibitem{Vog89}
K. Vogel and H. Risken,
Phys. Rev. A {\bf 40}, 2847 (1989).

\bibitem{Man97}
S. Mancini, V. I. Man'ko, and P. Tombesi,
Phys. Lett. A {\bf 213}, 1 (1996);\\
S. Mancini, V. I. Man'ko, and P. Tombesi,
Found. Phys. {\bf 27}, 801 (1997).

\bibitem{Kuh70}
T. S. Kuhn, {\it The Structure of Scientific Revolutions},
(University of Chicago Press, 1970).

\bibitem{Man99}
O.V. Man'ko and V.I. Man'ko, J.Russ Las.Res. {\bf 18}, 407  (1997).

\bibitem{Rad17}
J. Radon, Ber. Verh. S\"achs. Acad. {\bf 69}, 269 (1917).

\bibitem{DMM98}
V. V. Dodonov, O. V. Man'ko, V. I. Man'ko and A. W\" unsche,
Phys. Scripta {\bf 59}, 81 (1998).

\bibitem{OVM04}
O. V. Man'ko and V. I. Man'ko, arXiv:quant-ph/0407183;\\
V. I. Man'ko, G. Marmo, A. Simoni, A. Stern and F. Ventriglia,
Phys. Lett. A {\bf 343}, 251 (2005).

\bibitem{Gar}
C. W. Gardiner, {\it Quantum Noise}, (Springer, Berlin, 1991).

\bibitem{Pru77}
E. Prugovecki, Int. J. Th. Phys. {\bf 16}, 321 (1977).

\bibitem{BL}
P. Busch and P. J. Lahti, Ann. der Phys.  {\bf 47}, 369 (1990).

\bibitem{Dar00}
G. Cassinelli, G. M. D'Ariano,
E. De Vito and A. Levrero,
J. Math. Phys. \textbf{41}, 7940 (2000).

\bibitem{Fuc02}
C. Fuchs, arXiv:quant-ph/0205039.

\bibitem{OVM05}
O. V. Man'ko, V. I. Man'ko, G. Marmo, A. Shaji, E. C. G. Sudarshan and F. Zaccaria,
arXiv:quant-ph/0502089.

\bibitem{Evg}
S. Mancini, V. I. Man'ko, E. V. Shchukin and P. Tombesi,
J. Opt. B {\bf 5}, 333 (2003);\\
V. I. Man'ko, E. V. Shchukin and  S. Mancini, in preparation.

\bibitem{WZ}
J. A. Wheeler and W. Zurek, {\it Quantum Theory and Measurement}, (Princeton University Press, Princeton, 1983).

\bibitem{PR}
H. Primas,  {\it Chemistry, Quantum Mechanics, and Reductionism},
(Springer-Verlag, Berlin 1983).

\bibitem{WE}
A. Wehrl, Rev. Mod. Phys. {\bf 50}, 221 (1978). 

\bibitem{QLogic}
M. L. Dalla Chiara and R. Giuntini, {\it Quantum Logics},
(Kluwer, Amsterdam, 2002).

\bibitem{Mac}
G. Mackey, {\it The Mathematical Foundations of Quantum
  Mechanics}, (Benjamin, New York, 1957).

\bibitem{Bohr}
N. Bohr, Phys. Rev. {\bf 48}, 696 (1935).

\bibitem{OVM97}
V. I. Man'ko and O. V. Man'ko, JETP {\bf 85}, 430 (1997).

\bibitem{PLA}
R.A. Mosna and J. Vaz Jr., Phys. Lett. A {\bf 315}, 418 (2003)


\end{thebibliography}
\end{document}